\documentclass[12pt]{article}

\usepackage{graphicx}
\begin{document}
\begin{center}
{\Large Relativistic, Causal Description of Quantum Entanglement}
\vskip 0.3 true in
{\large J. W. Moffat}
\date{}
\vskip 0.3 true in
{\it Department of Physics, University of Toronto, Toronto, Ontario M5S1A7,
Canada,
\vskip 0.1 true in
and
\vskip 0.1 true in
Perimeter Institute for Theoretical Physics, Waterloo, Ontario N2J2W9,
Canada.}

\end{center}

\begin{abstract}%
A possible causal solution to the problem of providing a spacetime
description of the transmission of signals in quantum entangled
states is described using a `bimetric' spacetime structure, in which
the quantum entanglement measurements alter the structure of spacetime.
Possible experimental tests to verify the suggested spacetime
interpretation of quantum entangled states are proposed.

\end{abstract}
\vskip 0.2 true in e-mail: john.moffat@utoronto.ca

\vskip 0.3 true in

One of the most important features of quantum mechanics is the
Einstein-Podolsky-Rosen~\cite{Einstein,Schrodinger} effect, in which strong
correlations are observed between presently noninteracting particles that
have interacted in the past. The problem of understanding the consequences
of the EPR effect is still controversial~\cite{Laloe}. Experiments on {\it
entangled} particle states have verified the `nonlocal' nature of quantum
mechanics~\cite{Gisin}. One disturbing feature of the standard
interpretation of quantum mechanics is that the nonlocal nature of the
entanglement process has been divorced from our common intuitive ideas
about spacetime events and causality. The standard interpretation asserts
that for photons (or electrons) positioned at A and B, separated by a
spacelike distance, there is no exchange of classical information and
superluminal signals between A and B are impossible according to special
relativity. With the advent of the possibility of constructing quantum
computers and performing `teleportation' experiments, the whole issue of
the spacetime reality of the EPR process becomes more problematic.

In the following, we propose a scenario based on a `bimetric' description
of spacetime. This kind of construction has been successful in cosmology,
in which it provides an alternative to the standard inflationary
cosmologies~\cite{ClaytonMoffat}.
In the present application of the bimetric description of spacetime to the
quantum entanglement problem, we picture that a quantum mechanical metric
(QMM) frame is related to a special relativity metric (SRM) frame by the
gradients of a scalar field $\phi$. The SRM frame describes a Minkowski
light cone with constant speed of light $c_0$, while the QMM frame
describes a different light cone with an increased speed of light $c>c_0$.
The scalar field $\phi$ measures the amount of entanglement of a given
quantum mechanical state through the density matrix of its von Neumann
entropy. For a pure non-entangled state $\phi=0$, and
transmitted signals travel with the standard classical, special relativity
value $c_0$, but for entangled states, $\phi\not= 0$, quantum mechanical
superluminal signals can travel in the QMM frame, thereby providing a
Lorentz invariant {\it spacetime} description of quantum entanglement
phenomena. An observer in the SRM frame would see an apparent violation of
causality as a superluminal signal is transmitted between two entangled
sub-systems, while an observer in the QMM frame would see a causal
transmission of information between the two time-like separated
sub-systems.

We postulate that four-dimensional spacetime is described locally by the
bimetric structure:
\begin{equation}
\label{bimetric}
q_{\mu\nu}=\eta_{\mu\nu}+\beta\partial_\mu\phi\partial_\nu\phi,
\end{equation}
where $\beta>0$ is a constant with dimensions of $[{\rm length}]^2$ and we choose the
scalar field $\phi$ to be dimensionless. The metric $q_{\mu\nu}$ is called the
quantum mechanical metric (QMM), while $\eta_{\mu\nu}$ is the
special relativity metric (SRM) described by the Minkowski metric $\eta_{\mu\nu}={\rm
diag}(1,-1,-1,-1)$. The inverse metric $q^{\mu\nu}$ satisfies
\begin{equation}
q^{\mu\alpha}q_{\nu\alpha}={\delta^\mu}_\nu.
\end{equation}
We assume that
only non-degenerate values of $q_{\mu\nu}$ with ${\rm
Det}(q_{\mu\nu})\not= 0$ correspond to physical spacetime. The special
relativity metric is given by
\begin{equation}
\label{SRM}
ds^2\equiv\eta_{\mu\nu}dx^\mu dx^\nu=c_0^2dt^2-(dx^i)^2,
\end{equation}
where ($i,j=1,2,3$). The quantum mechanical metric is
\begin{equation}
\label{QQM}
ds_q^2\equiv q_{\mu\nu}dx^\mu dx^\nu=(\eta_{\mu\nu}+\beta\partial_\mu\phi\partial_\nu\phi)dx^\mu dx^\nu.
\end{equation}
\label{QMM2}
The latter can be written as
\begin{equation}
ds_q^2=c^2_0\biggl(1+\frac{\beta}{c_0^2}{\dot\phi}^2\biggr)dt^2-(\delta_{ij}-\beta\partial_i\phi
\partial_j\phi)dx^idx^j,
\end{equation}
where $\dot\phi=d\phi/dt$. We see that the speed of light in the QMM frame is time
dependent:
\begin{equation}
c(t)=c_0\biggl(1+\frac{\beta}{c_0^2}\dot\phi^2\biggr)^{1/2}.
\end{equation}
If we choose $\dot\phi\approx 0$, then the speed of light has a
purely spatial dependence given by
\begin{equation}
c({\vec x})\equiv\vert d{\vec x}/dt\vert=\frac{c_0}{[1-\beta(\partial_i\phi)^2]^{1/2}},
\end{equation}
and we require that $\beta(\partial_i\phi)^2<1$.

The null cone equation $ds^2=0$ describes light signals
moving with the constant measured speed $c_0$, whereas $ds_q^2$ cannot be
zero along the same null cone lines, but vanishes for an expanded null
cone with the speed of light $c > c_0$.

Both metrics (\ref{SRM}) and (\ref{QQM}) are
invariant under Lorentz transformations
\begin{equation} x^{'\mu}={\Lambda^\mu}_\nu x^\nu,
\end{equation}
where ${\Lambda^\mu}_\nu$ are constant matrix coeeficients which satisfy the
orthogonality condition
\begin{equation}
{\Lambda^\mu}_\nu{\Lambda_\mu}^\sigma={\delta_\nu}^\sigma.
\end{equation}

We shall require a relativistic description of quantum mechanics. To this
end, we introduce the concept of a general space-like surface in
the Minkowski spacetime of the SRM frame, instead of the flat surface
$t={\rm constant}$. We demand that the normal to the surface at any point
$x$, $n_\mu(x)$ be time-like: $n_\mu(x)n^\mu(x) > 0$. We shall denote a
spacelike surface by $\sigma$. A local time $t({\vec x})$ is assigned, so
that in the limit that the surface becomes plane, each point has the same
time $t={\rm constant}$. We can now define the Lorentz invariant
functional derivative operation $\delta/\delta\sigma(x)$ and  the
Tomonaga-Schwinger equation ~\cite{Tomonager,Schwinger,Schweber}
\begin{equation}
\label{Tomonaga-Schwinger}
i\hbar c_0\frac{\delta\psi(\sigma)}{\delta\sigma(x)}={\cal
H}(x)\psi(\sigma), \end{equation} where
\begin{equation}
H(t)=\int d^3x{\cal H}(x)
\end{equation}
is the Hamiltonian operator and ${\cal H}(x)$ is the Lorentz invariant
Hamiltonian density. Eq. (\ref{Tomonaga-Schwinger}) is a relativistic
extension of the Schr\"odinger equation
\begin{equation}
i\hbar\partial_t\psi(t)=H(t)\psi.
\end{equation}

The domain of variation of $\sigma$ is restricted by
the integrability condition
\begin{equation}
\label{comm}
[{\cal H}(x),{\cal H}(x')]=0,
\end{equation}
for $x$ and $x'$ on the space-like surface $\sigma$. In quantum field
theory, it is usual to work in the interaction picture, so that the
invariant commutation rules for the field operators automatically guarantee
that (\ref{comm}) is satisfied for all interacting fields with local
nonderivative couplings.

We introduce the concept of a Lorentz invariant density matrix:
\begin{equation}
\rho[\sigma(x)]=\vert\psi(\sigma)\rangle\langle\psi(\sigma)\vert.
\end{equation}
In the Heisenberg representation
\begin{equation}
\rho(\sigma)=\exp(-i{\cal H}(x)\sigma/\hbar c_0)\rho(0)\exp(i{\cal H}(x)\sigma/\hbar c_0).
\end{equation}
Then, the density matrix operator $\rho(\sigma)$ satisfies the invariant
Heisenberg equations of motion
\begin{equation}
i\hbar c_0\frac{\delta\rho[\sigma(x)]}{\delta\sigma(x)}
=[{\cal H}(x),\rho(\sigma)].
\end{equation}

We will relate the `quantum information' scalar field $\phi$ to a measure of
the entanglement of a pure quantum state by the relativistic definition
\begin{equation}
\phi[\sigma(x)]= \gamma S(\rho_A)[\sigma(x)]=\gamma S(\rho_B)[\sigma(x)],
\end{equation}
where $\gamma$ is a constant and
\begin{equation}
\label{relaentropy}
S(\rho_m)[\sigma(x)]=-{\rm Tr}_m\rho[\sigma(x)]\log\rho[\sigma(x)]
\end{equation}
is the relativistic scalar entropy of the sub-systems A and B. Moreover,
$\rho_A={\rm Tr}_B\vert\psi\rangle\langle\psi\vert$ is the reduced density
matrix obtained by tracing the whole system's pure state density matrix
$\rho_{AB}=\vert\psi\rangle\langle\psi\vert$ over A's degrees of freedom,
while $\rho_B={\rm Tr}_A\vert\psi\rangle\langle\psi\vert$ is the partial
trace over B's degrees of freedom. For a non-entangled state for which the
$\psi(\sigma)$ can be expressed as a tensor product
$\psi_A(\sigma)\otimes\psi_B(\sigma)$, we have $\phi(\sigma)=0$. In the
non-relativistic limit, (\ref{relaentropy}) reduces to the pure bi-particle
entropy measure of entanglement~\cite{Bennett,Wootters}. For the scalar field
$\phi$, we have
\begin{equation}
\partial_\mu\phi[\sigma(x)]=\frac{\delta\phi[\sigma(x)]}{\delta\sigma(x)}
\int_{\sigma}\phi(x')d\sigma'_{\mu}.
\end{equation}

We consider in the following only the simple physical system of two photons
(or two electrons). The entanglement measure for a mixed state and a
multiparticle state is controversial and no consensus has been reached on
how to define it.

We postulate that the scalar field $\phi$ satisfies the equation of motion
\begin{equation}
\label{scalarmotion}
\partial_\mu\partial^\mu\phi+V'(\phi)=0,
\end{equation}
where $V(\phi)$ is the potential for the field $\phi$ and $V'(\phi)
=\partial V(\phi)/\partial\phi$. For a free scalar field $\phi$ the potential will be
$V(\phi)=\frac{1}{2}m^2\phi^2$, where $m$ is the mass of the particle
associated with the scalar field $\phi$.

For a pure non-entangled state $\phi=0$ the spacetime metric is
described by (\ref{SRM}), whereas for an entangled state $\phi\not= 0$
and $\partial_\mu\phi\not=0$ the spacetime is described by (\ref{QQM}). Since the speed of light
can become much larger in the QMM spacetime frame, it is now possible to
transmit signals at `superluminal' speeds without violating the causality
notions that prevail in the familiar SRM spacetime frame. In this way, by
introducing a bimetric spacetime, we have incorporated the notions of
spacetime events and causality for quantum mechanical entangled states.
For an observer who detects a quantum mechanical system with some
measuring device and observes a pure non-entangled state, the laws of
special relativity in his SRM frame are consistent with the finite
measured speed of light $c_0$, while an observer who makes a measurement
of an entangled quantum mechanical system
will cause $\partial_\mu\phi$ to be non-zero, and he will observe in the QMM
frame an exchange of information with the spatially distant other
component of the entangled state.  In the QMM frame the speed of the light
signal $c$ emitted, say, at a counter at A and received at a distant
counter B will be {\it finite} but large compared to the measured value of
the speed of light $c_0$ in the SRM frame, due to the expanded light cone
in the QMM frame. In the QMM frame, for a maximal entanglement, A and B
may no longer be `spacelike' separated.

When $\phi={\rm constant}$, the speed of light is constant in
space and time and a constant entanglement would be associated with the
SRM frame. However, it is to be expected that $S(\rho_m)$ varies in space
and time when a measurement is performed on the entangled state and,
therefore, in practice $\phi$ will be spacetime dependent.

We have considered the possibility that state reduction is a well-defined
physical process that leads to a localized spacetime causality in the QMM
spacetime frame~\cite{Kent}. Before the state reduction, the relativistic
Schr\"odinger equation (\ref{Tomonaga-Schwinger}) evolves unitarily in the
SRM frame.

The current correlation experiments have a separation distance $\sim$ 11
kilometres~\cite{Gisin}. The results of experiments performed in
Geneva~\cite{Zbinden} have been analyzed in a preferred frame
formalism~\cite{Caban}, choosing the cosmic microwave background
radiation as the frame~\cite{Scarani}. A lower bound for the speed of
superluminal quantum information in this frame is $1.5\times 10^4c_0$.
Thus, for a fixed time $t$ in the QMM frame, a superluminal speed of light
signal of order $10^4-10^7c_0$ would be consistent with a finite, causal
exchange of information between the two sub-systems A and B.

A possible experiment to verify the bimetric spacetime scenario would be to attempt
to measure the space and time dependence of the entanglement field $\phi$. If we
assume that the solution of the equation of motion (\ref{scalarmotion}) for $\phi$,
given a potential $V(\phi)$, falls off with distance and $V(\phi)$ vanishes
asymptotically at infinity, then for sufficiently spatially separated
sub-systems A and B, we would observe a decrease in the amount of
entanglement of the quantum states at A and B and the measured correlation
would decrease. Thus, the originally prepared entangled states would show a
decreasing violation of Bell's inequalities~\cite{Bell}.

Another possibility to experimentally verify our interpretation of quantum
mechanics is when the entanglement field $\phi$ interacts with matter through the
potential $V(\phi)$. If A and B are spatially separated with a large intervening
mass, then, for a suitable potential $V(\phi)$, the interaction between
$\phi$ and the matter could decrease $\phi$, thereby diminishing the
measured correlation, when compared to the same experimental setup
without the intervening mass.

Provided that the diminished correlation
signature can be distinguished from potential decoherence effects, the
consequences of our proposed interpretation of quantum mechanics can be
experimentally tested. In the standard picture of quantum mechanics there
is, apart from potential decoherence effects, no dynamical spatial or time
dependence predicted for the correlation between two entangled states A
and B.

We can generalize our theory to include gravitation by replacing
$\eta_{\mu\nu}$ in (\ref{bimetric}) by $g_{\mu\nu}$, the pseudo-Riemannian
metric of spacetime. Then, the speed of light waves and
gravitational waves will both be dynamical and vary in spacetime along the
light cone in the QMM frame determined by $ds_q^2=0$. Before a measurement
is made of the entangled system and the ensuing state reduction occurs, the
classical action for the gravitational theory will be given by
\begin{equation}
S=S_g+S_\phi+S_M,
\end{equation}
where
\begin{equation}
S_g[g]=-\frac{1}{\kappa}\int d^4x\sqrt{-g}(R[g]+2\Lambda),
\end{equation}
and
\begin{equation}
S_\phi[\phi,g]=\frac{1}{\kappa}\int
d^4x\sqrt{-g}\biggl(\frac{1}{2}g^{\mu\nu}\partial_\mu\phi\partial_\nu\phi-V(\phi)\biggr).
\end{equation}
Moreover, the matter stress-energy tensor is
\begin{equation}
T_{\mu\nu}=-\frac{2}{\sqrt{-g}}\frac{\delta S_M}{\delta g^{\mu\nu}},
\end{equation}
$\kappa=16\pi G/c_0^4$, and $\Lambda$ is the cosmological constant. When a
quantum entangled state is detected and
$\partial_\mu\phi\not= 0$, then the metric $g_{\mu\nu}$ is replaced by
\begin{equation}
g_{\mu\nu}=q_{\mu\nu}-\beta\partial_\mu\phi\partial_\nu\phi.
\end{equation}

The consequences that follow from an application of the gravitational QMM
theory will be considered elsewhere.

\vskip 0.3 true in

{\bf Acknowledgment}
\vskip 0.2 true in
This work was supported by the Natural Sciences and Engineering Research Council of
Canada. I thank Dr. Valerio Scarani for providing me with helpful
information.

\vskip 0.5 true in

\end{document}